\documentclass[10pt]{article}
\usepackage{graphicx}
\usepackage{fullpage}
\usepackage{amsmath}
\usepackage{epsfig}
\usepackage{amssymb, graphicx}
\usepackage{wrapfig}
\begin{document}
\title{Raman scattering study of delafossite magnetoelectric multiferroic compounds:\\ CuFeO$_2$ and CuCrO$_2$}
\author{ Oktay Aktas, Kim Doan Truong, Tsuyoshi Otani, Geetha Balakrishnan,\\ Maynard J. Clouter, Tsuyoshi Kimura, and Guy Quirion}
\date{\today}
\maketitle

\begin{abstract}
Ultrasonic velocity measurements on the magnetoelectric multiferroic compound CuFeO$_2$ reveal that the antiferromagnetic transition observed at $T_{N1}$ = 14 K might be induced by an $R\overline{3}m \rightharpoonup C2/m$ pseudoproper ferroelastic transition \cite{quirion2008}. In that case, the group theory states that the order parameter associated with the structural transition must belong to a two dimensional irreducible representation E$_g$ ($x^2 - y^2$, $xy$). Since this type of transition can be driven by a Raman E$_g$ mode, we performed Raman scattering measurements on CuFeO$_2$ between 5 K and 290 K. Considering that the isostructural multiferroic compound CuCrO$_2$ might show similar structural deformations at the antiferromagnetic transition $T_{N1}$ = 24.3 K, Raman measurements have also been performed for comparison. At ambient temperature, the Raman modes in CuFeO$_2$ are observed at $\omega_{\mathrm{E}_g}$ = 352 cm$^{-1}$ and $\omega_{\mathrm{A}_g}$ = 692 cm$^{-1}$, while these modes are detected at $\omega_{\mathrm{E}_g}$ = 457 cm$^{-1}$ and $\omega_{\mathrm{A}_g}$ = 709 cm$^{-1}$ in CuCrO$_2$. The analysis of the temperature dependence of modes shows that the frequency of all modes increases down to 5 K. This typical behavior can be attributed to anharmonic phonon-phonon interactions. These results clearly indicate that none of the Raman active modes observed in CuFeO$_2$ and CuCrO$_2$ drive the pseudoproper ferroelastic transition observed at the N\'{e}el temperature $T_{N1}$. Finally, a broad band at about 550 cm$^{-1}$ observed in the magnetoelectric phase of CuCrO$_2$ below $T_{N2}$ could be attributed to a magnon mode.

\end{abstract}

\section{Introduction}

CuFeO$_2$ and CuCrO$_2$ belong to the delafossite frustrated antiferromagnets with the chemical formula ABO$_2$ in which A is a nonmagnetic monovalent ion (Cu and Ag) while B is a magnetic trivalent ion such as Fe and Cr \cite{kimura2006,seki2008, quirion2009,kimura2008}. Some of these compounds, including AgCrO$_2$, CuFeO$_2$, and CuCrO$_2$, belong to the trigonal $R\overline{3}m$ space group at room temperature and undergo a series of magnetic phase transitions \cite{seki2008, quirion2009, kimura2008} at low temperatures as a result of geometrical frustration of magnetic ions sitting on a triangular lattice. 

In the case of CuFeO$_2$, two antiferromagnetic transitions are observed at zero field. In its ground state, Fe$^{+3}$ ions order into a collinear commensurate four-sublattice ($ \uparrow \uparrow \downarrow \downarrow $) structure, while between $T_{N2}$ = 11 K and $T_{N1}$ =~14 K, the magnetic order is incommensurate with the magnetic moments also pointing along the c-axis \cite{mitsuda1998}. With the application of a field parallel to the c-axis, a series of new magnetic orders is stabilized below T$_{N2}$. Between 7 T and 13 T, CuFeO$_2$ shows a proper screw spin configuration where the spins lie in the $R\overline{3}m$ mirror plane perpendicular to the magnetic modulation vector $\mathbf{q} \| [110] $ (hexagonal basis) \cite{kimura2006, quirion2009, terada2006, ajiro1994}. At higher fields, several other spin configurations are observed: a c-axis collinear 5-sublattice ($\uparrow \uparrow \uparrow \downarrow \downarrow $) state (13 T $<$ H $<$ 20 T), a c-axis collinear 3-sublattice ($\uparrow \uparrow \downarrow$) structure (20 T $<$ H $<$ 34 T), a canted 3-sublattice state (34 T $<$ H $<$ 49 T), and a noncolinear incommensurate spin-flop phase which is close to the 120$^{\circ}$ spin structure for 49 T $<$ H $<$ 70 T, followed by a transition to the paramagnetic state at 70 T. \cite{quirion2009, terada2007}.

While CuCrO$_2$ is isostructural to CuFeO$_2$ at room temperature, its magnetic phase diagram is significantly different \cite{seki2008, kimura2008, kadowakit}. According to specific heat and magnetic susceptibility measurements \cite{kimura2008}, CuCrO$_2$ shows anomalies at $T_{N1}$ = 24.3 K and $T_{N2}$ = 23.6 K. The magnetic order in the intermediate temperature range $T_{N1} < T < T_{N2}$ is interpreted as a collinear state with \textbf{S}$\| c $ \cite{kimura2008}, while recent neutron diffraction measurements \cite{poienar2009, soda2009} reveal an incommensurate proper screw spin structure with $\mathbf{q} \| [110]$  below $T_{N2}$. This spin configuration is very similar to the one observed in CuFeO$_2$ between 7 T and 13 T. Moreover, additional studies on both compounds \cite{kimura2006, seki2008, kimura2008} show that an electric polarization $\mathbf{P} \| [110] $ is only induced upon the emergence of this proper-screw spin order. Under this scenario, the usual inverse Dzyaloshinskii-Moriya (DM) interaction $\mathbf{P} \sim \mathbf{r}_{ij} \times (\mathbf{S}_i \times \mathbf{S}_j)$ \cite{katsura2005, mostovoy2006} cannot account for the induced polarization as the q-vector of the spin modulation is perpendicular to the spiral-plane. An alternative possibility, proposed by Arima et al. \cite{arima2007}, is that the polarization is induced by spin-orbit coupling. Thus, CuFeO$_2$ and CuCrO$_2$ represent a different class of magnetoelectric multiferroics in which the mechanism leading to the magnetoelectric coupling is still uncertain.

\begin{figure}[htbp]
\begin{center}
		\includegraphics[width=12cm]{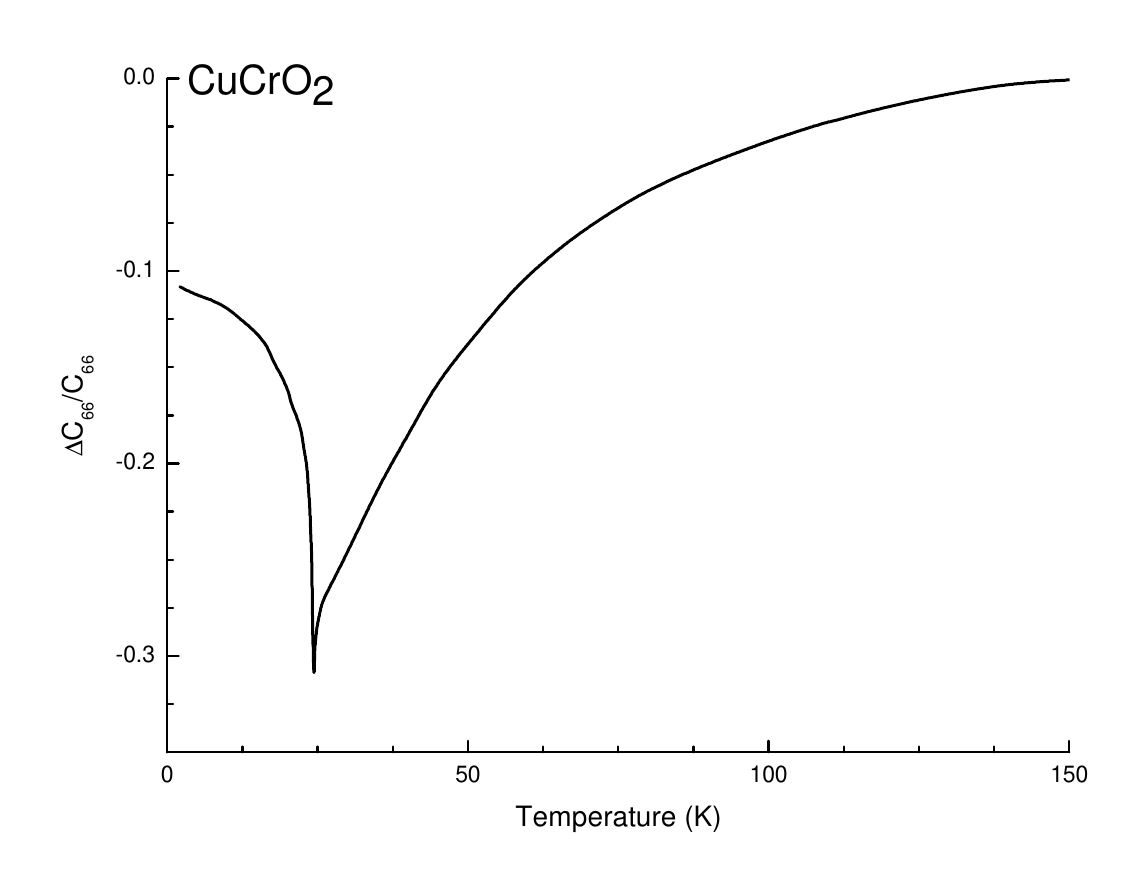}
		%\end{center}
		\caption{Relative variation of the elastic constant $C_{66}$ in CuCrO$_2$ as a function of temperature obtained by sound velocity measurements. $C_{66}$ shows a 30 $\%$ reduction at $T_{N1}$ = 24.3 K relative to the value at 150 K, indicating that the antiferromagnetic transition at $T_{N1}$ might also be ferroelastic. }
		\label{C66}
		\end{center}
		\end{figure}

Other particular properties of CuFeO$_2$ have also been recently revealed via sound velocity measurements \cite{quirion2008,quirion2009, quirion2008b}. These measurements show softening on specific elastic constants as the temperature is decreased down to $T_{N1}$ = 14 K. The data analysis indicates that this peculiar behavior is characteristic of an $R\overline{3}m \rightharpoonup C2/m $ pseudoproper ferroelastic transition, consistent with neutron \cite{ye2006} and x-ray \cite{terada2006} diffraction measurements. Furthermore, according to the group theory \cite{tinkhan}, the order parameter associated with the structural transition must belong to a two dimensional irreducible representation (IR) E$_g$ ($x^2 - y^2$, $xy$). As none of the spin components belong to this IR, these measurements indicate that the magnetic order in CuFeO$_2$ is stabilized by the ferroelastic structural transition, putting in evidence the role played by the spin-lattice coupling in this family of multiferroic materials. Thus, the true origin of the structural transition observed at $T_{N1}$ remains a mystery. One possibility is that the transition is driven by a Raman mode as in other pseudoproper ferroelastic materials \cite{oktay, quirion2009RLHS, sno2, bivo4}. Regarding isostructural CuCrO$_2$, recent megnetostriction measurements \cite{kkimura2009} show evidence for a structural phase transition at $T_{N1}$ = 24.3 K. Furthermore, as preliminary sound velocity measurements on CuCrO$_2$ (shown in Fig.~\ref{C66}) reveal softening similar to that observed in CuFeO2 \cite{quirion2008}, the transition observed at $T_{N1}$ = 24.3 K in CuCrO2 might also be ferroelastic. In order to possibly identify the order parameter associated with these pseudoproper ferroelastic transitions, we performed Raman scattering measurements on the isostructural compounds CuFeO$_2$ and CuCrO$_2$.
The remainder of the paper is organized as follows. Experimental methods are discussed in Sec.~\ref{experiment} while results and discussion are presented in Sec.~\ref{results}. Finally, conclusions are made in Sec.~\ref{conclusions}

\section{Experiment \label{experiment}}
 Single crystals of CuFeO$_2$ were grown by the floating zone method using a four mirror image furnace \cite{petrenko2000}. CuFeO$_2$ samples used in the measurements had an area of $\sim$2 mm x $\sim$2 mm and was $\sim$1 mm long along the c-axis. Single crystals of CuCrO$_2$ were grown from Bi$_2$O$_3$ flux \cite{kimura2008}. The samples were platelets with a length of 0.4 mm along the c-axis. The surface area was approximately 2 mm x 2 mm. Prior to Raman scattering experiments, samples were polished using abrasive slurry with 50 nm Al$_2$O$_3$ grains in order to minimize surface scattering. Room temperature Raman measurements were performed using two different experimental setups. Using an Ar$^{+}$ laser operating at 514.5 nm, the Raman spectra were collected by a double grating spectrometer (Spex Industries, model 1401), a photomultiplier tube (Perkin Elmer, MP 900 series), and a photon counter (Princeton Applied Research, model 1109). For CuCrO$_2$, 28 mW of exciting beam power was used while it was increased to 50 mW to obtain the spectra of CuFeO$_2$. For cross polarization measurements on CuFeO$_2$, the beam power was increased up to 100 mW which caused local heating on the sample. In the second setup, the 0.5 cm$^{-1}$ resolution micro-Raman measurements were performed with a 632 nm He-Ne laser, a double grating spectrometer (Jobin Yvon, model Labram-800) and a liquid-nitrogen cooled CCD detector. In order to minimize sample heating, 0.3 mW incident beam power with a 3 $\mu$m spot size (4000 W/cm$^{2}$) was used. Mode parameters were obtained by a fit to the data using Lorentzian functions for the observed modes.

\section{Experimental results and discussion \label{results}}
\subsection{Raman spectra of CuFeO$_2$ and CuCrO$_2$ at room temperature \label{rt}}

\begin{figure}[htbp]
\begin{center}
		\includegraphics[width=12cm]{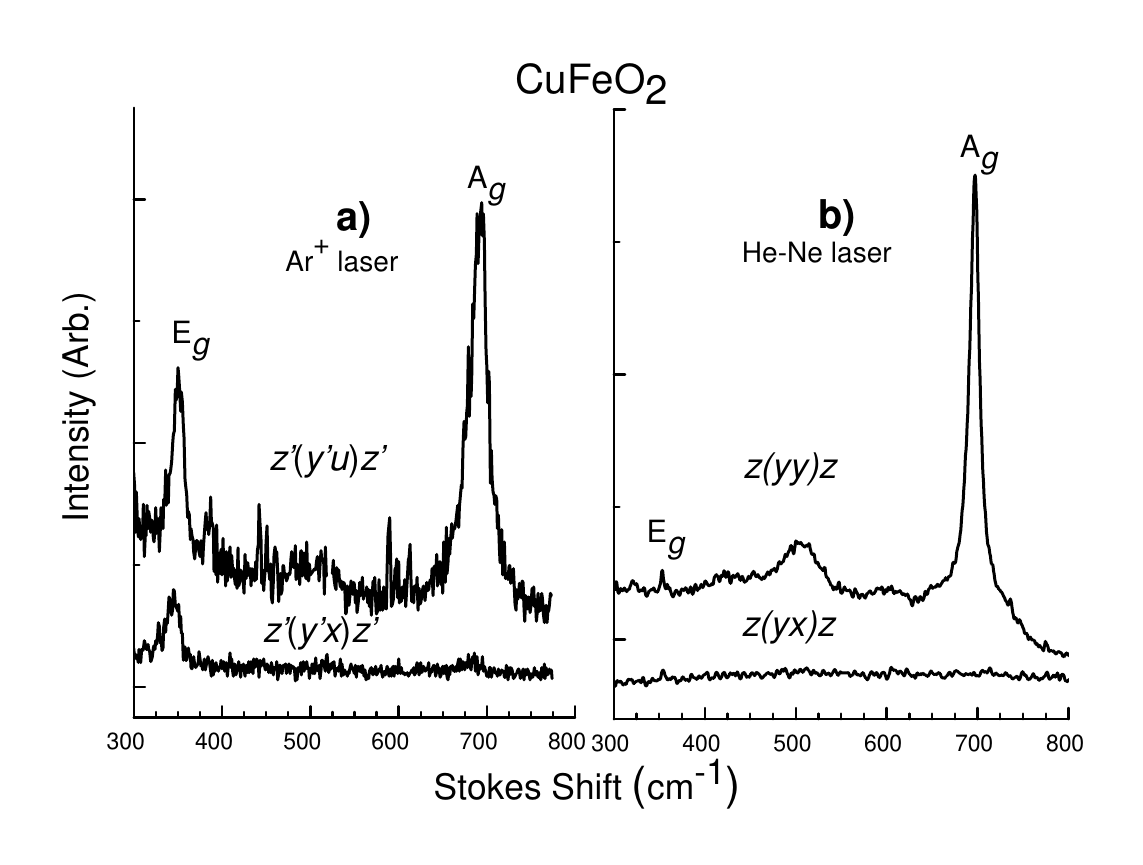}
		%\end{center}
		\caption{Polarized Raman Spectra of CuFeO$_2$ at room temperature obtained using the Ar$^{+}$ and He-Ne lasers. Experimental scattering geometries are represented by the Porto notation above each spectrum. Polarizations along the $y'$ and $x$ axes are parallel and perpendicular to the plane of incidence, respectively. Strong plasma line at 521 cm$^{-1}$ in the $z'(y'u)z'$ spectrum was removed for clarity. Raman modes with A$_g$ and E$_g$ symmetries are located at $\omega_{\mathrm{A}_g}$ = 692 cm$^{-1}$ and $\omega_{\mathrm{E}_g}$ = 351 cm$^{-1}$, respectively. }
		\label{ramanRT}
		\end{center}
		\end{figure}

\begin{figure}[htbp]
\begin{center}
		\includegraphics[width=12.0cm]{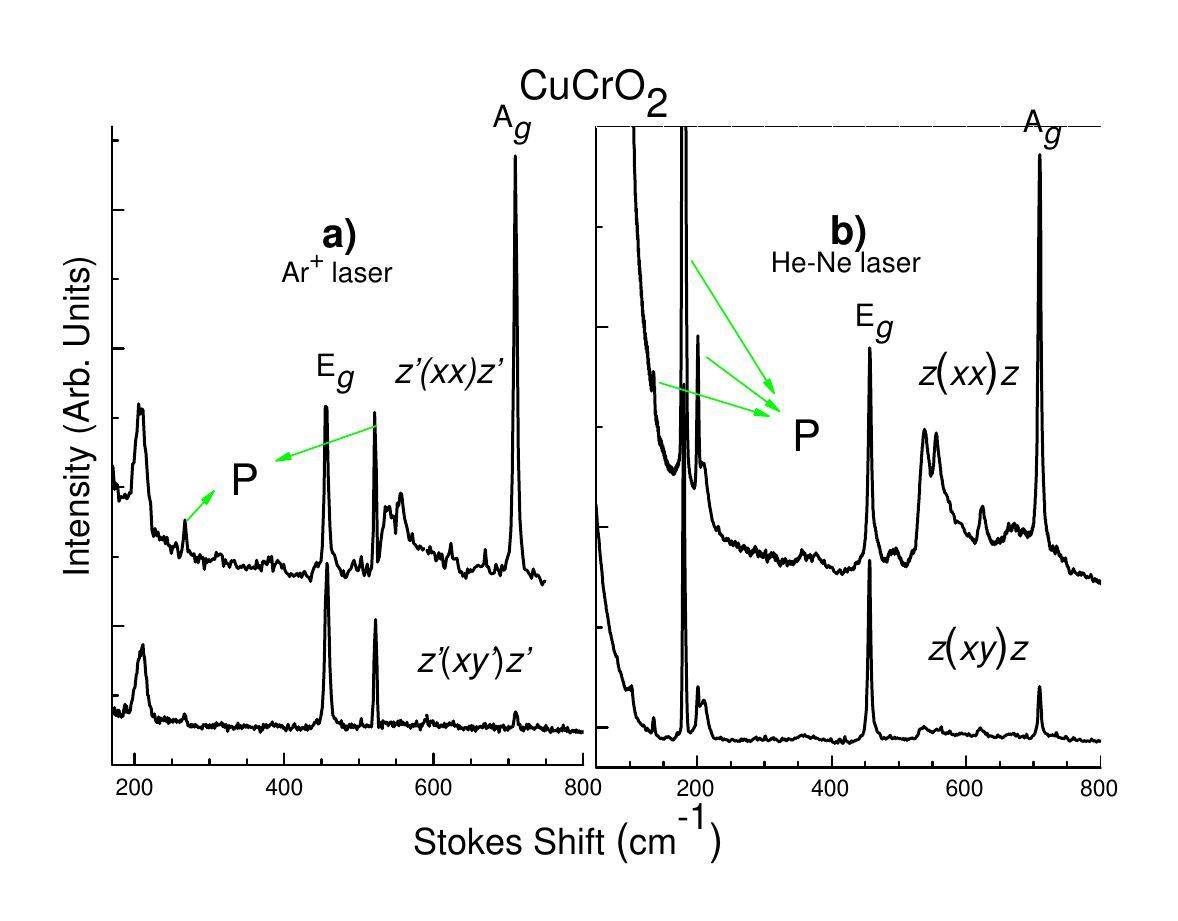}
		%\end{center}
		\caption{Polarized Raman Spectra of CuCrO$_{2}$ at room temperature obtained using the Ar$^{+}$ and He-Ne lasers. Experimental scattering geometries are designated by the Porton notation above each spectrum. Polarizations along the $y'$ and $x$ axes are parallel and perpendicular to the plane of incidence, respectively. Green arrows indicate the plasma lines (P). Raman modes have frequencies at $\omega_{\mathrm{E}_g}$ = 457 cm$^{-1}$ and $\omega_{\mathrm{A}_g}$ = 709 cm$^{-1}$.}
		\label{ramanrt1}
		\end{center}
		\end{figure}

Delafossite compounds (space group $R\overline{3}m$) such as CuFeO$_2$ and CuCrO$_2$ have one formula unit per unit cell with a total of 12 possible vibrational modes. Among these modes only two are Raman active with E$_g$ and A$_g$ symmetry. The A$_g$ mode corresponds to vibrations of the Cu-O bonds along the c-axis while the E$_g$ mode represents vibrations in the triangular lattice perpendicular to the c-axis. The atomic displacements for these modes are illustrated in Ref. \cite{porres2005}. In order to determine the symmetry of the modes observed in CuFeO$_2$ (Fig.~\ref{ramanRT}) and CuCrO$_2$ (Fig.~\ref{ramanrt1}), we performed polarized Raman scattering measurements at room temperature using two different laser sources. Here, Raman scattering geometries are identified using the Porto notation $k_i(e_i e_s)k_s$. The labels $z'$ and $y'$ designate directions making an angle $\theta$ relative to the z and y axes, where $\theta = 50^\circ$ for CuFeO$_2$ while $\theta = 15^\circ$ in the case of CuCrO$_2$.

To our knowledge, no polarized Raman measurements on CuFeO$_2$ single crystals have been reported. At room temperature, the spectrum taken with the Ar$^+$ laser using unpolarized ($u$) scattered light, Fig. \ref{ramanRT}a, shows modes at 349 cm$^{-1}$ and 690 cm$^{-1}$ in good agreement with results obtained on polycrystals \cite{pavunny2009, pavunny2010}. The intensity of the mode at 690 cm$^{-1}$ disappears with cross ($y'x$) polarization while the mode at 349 cm$^{-1}$ remains visible in both $(y'u)$ and cross ($y'x$)  polarizations. Measurements with the He-Ne laser show Raman modes at 351 cm$^{-1}$ and 692 cm$^{-1}$ and a broad band at 496 cm$^{-1}$ (See Fig. \ref{ramanRT}b). The mode at 692 cm$^{-1}$ has a strong intensity in the parallel polarization ($yy$) and disappears in the cross  polarization ($yx$). The intensity of the mode at 351 cm$^{-1}$ is very weak which implies that the He-Ne excitation line at 632.8 nm is not in resonance with the vibrations associated with this mode as observed in LiNiO$_2$ \cite{julien2002}. Despite its weak intensity, it is visible in both polarizations. Moreover, this mode was reproducible down to low temperatures (see Fig.~\ref{ccocfot}b). According to the Raman scattering tensors associated with the trigonal point group $\overline{3}m$ \cite{decius},
\begin{equation}
\mathrm{A_g(x)} = \left[  \begin{array}{cccc}
a & 0 & 0 \\
0 & a& 0   \\ 
0 & 0 & b 
\end{array} \right]
\label{Agsym}
\end{equation}
and
\begin{equation}
\mathrm{E_g(x)} = \left[  \begin{array}{cccc}
c & 0 & 0 \\
0 & -c& d   \\ 
0 & d & 0 
\end{array} \right],\  \mathrm{E_g(y)} =  \left[  \begin{array}{cccc}
0 & -c & -d \\
-c & 0 & 0   \\ 
-d & 0 & 0 
\end{array} \right],
\label{Egsym}
\end{equation}
a cross polarization configuration such as $z(yx)z$ allows only $E_g$ modes, while a parallel polarization configuration like $z(xx)z$ allows the observation of $E_g$ and $A_g$ modes. Therefore, the mode symmetry is assigned as $\omega_{\mathrm{A}_g}$ = 692 cm$^{-1}$ and $\omega_{\mathrm{E}_g}$ = 351 cm$^{-1}$.

In addition to the vibrational modes observed in CuFeO$_2$, a broad band located at 496 cm$^{-1}$ is also revealed using both laser sources. In the unpolarized ($y'u$) spectrum obtained with the Ar$^+$ laser, the intensity of this feature is within the background noise. In the parallel polarized spectrum obtained with the He-Ne laser, the broad peak is clearly observed and detected down to 5 K (Fig.~\ref{ccocfot}). Polarized spectra obtained with both excitation lines show that the mode at 496 cm$^{-1}$ has an $A_g$ symmetry. The possible origin of this band will be discussed later.

As in the case of CuFeO$_2$, CuCrO$_2$ should show two Raman modes. However, with unpolarized ($u$) scattered light (not shown) or a parallel polarization ($xx$) configuration with the Ar$^+$ laser (Fig.~\ref{ramanrt1}a), we observe modes at 104 cm$^{-1}$, 207 cm$^{-1}$, 382 cm$^{-1}$ 457 cm$^{-1}$, 538 cm$^{-1}$, 557 cm$^{-1}$, 623 cm$^{-1}$, 668 cm$^{-1}$,  and 709 cm$^{-1}$. Using the He-Ne laser (Fig.~\ref{ramanrt1}b), a similar spectrum is obtained except for a mode at 382 cm$^{-1}$ using the Ar$^+$ laser and a mode at 359 cm$^{-1}$ in the case of the He-Ne laser. The symmetries of these modes can be assigned according to the polarized Raman measurements shown in Fig.~\ref{ramanrt1}. Since the modes at 104 cm$^{-1}$, 212 cm$^{-1}$, and 457 cm$^{-1}$ are observed in both parallel and cross polarized Raman spectra, these modes belong to the $E_g$ irreducible representation (IR). The other modes are therefore assigned to $A_g$ IR since their intensities either are weak or disappear in the cross polarization configuration. So far, there have been four publications reporting Raman spectra on CuCrO$_2$ powder samples \cite{ramancco1,ramancco4, ramancco2,ramancco3}. Two of these publications show modes at 207 cm$^{-1}$, 444 cm$^{-1}$, and 691 cm$^{-1}$ \cite{ramancco1, ramancco4}. In addition, one of these works shows additional features with weak intensities at 540 cm$^{-1}$ and 560 cm$^{-1}$ \cite{ramancco4}. Other publications \cite{ramancco2, ramancco3} reveal Raman modes only at 452 cm$^{-1}$ and 703 cm$^{-1}$. By comparison, our polarized Raman results indicate that the Raman modes in CuCrO$_2$ correspond to $\omega_{\mathrm{A}_g}$ = 709 cm$^{-1}$ and $\omega_{\mathrm{E}_g}$ = 457 cm$^{-1}$. 
		
As mentioned earlier, CuFeO$_2$ and CuCrO$_2$ should only have two Raman modes. However, both compounds show additional features (Figs. \ref{ramanRT} and \ref{ramanrt1}) similar to those observed in other delafossite compounds such as CuAlO$_2$ \cite{porres2006} and CuGaO$_2$ \cite{porres2005}. In agreement with \textit{ab initio} calculations, these additional modes in CuAlO$_2$ are attributed to non-zero wavevector phonons which are normally forbidden by Raman selection rules \cite{porres2006}. As suggested, the selection rules are possibly relaxed by defects such as Cu vacancies, interstitial oxygens or tetrahedrally coordinated Cr$^{+3}$ or Fe$^{+3}$ on the Cu site \cite{porres2006}. Thus, the additional features observed in CuFeO$_2$ and CuCrO$_2$ could have an origin similar to that observed in CuAlO$_2$ \cite{porres2006} and CuGaO$_2$ \cite{porres2005}. They could also be related to crystal field excitations which were revealed in Raman spectra of other geometrically frustrated compounds \cite{lummen2008}.

\subsection{Temperature dependent measurements \label{lt}}

\begin{figure}[htbp]
\begin{center}
		\includegraphics[width=12cm]{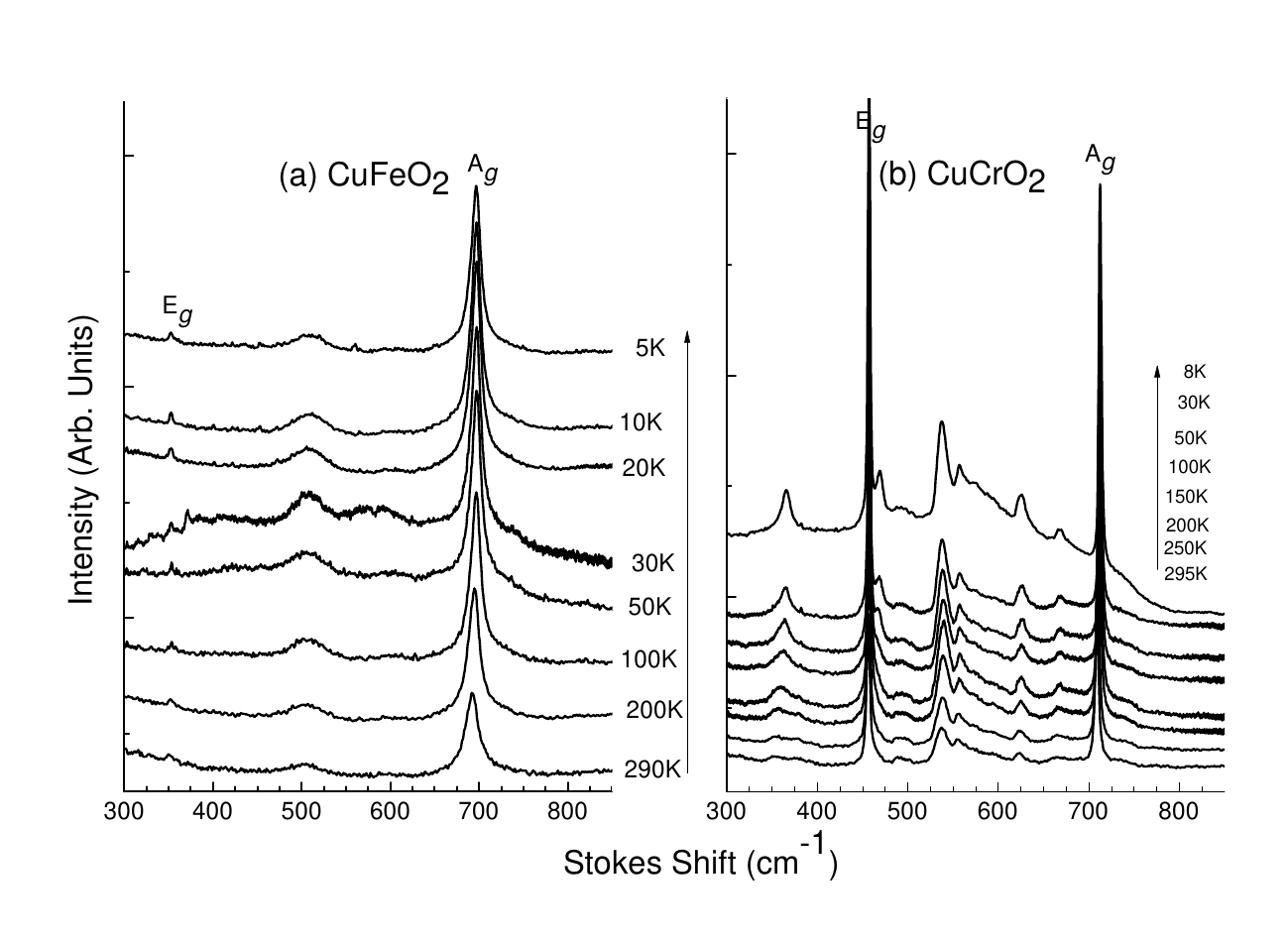}
		%\end{center}
		\caption{Raman spectra of (a) CuCrO$_2$ and (b) CuFeO$_2$ between 5 K and 290 K. While no considerable change is observed in the CuFeO$_2$ spectra in this temperature range, an additional mode in CuCrO$_2$ appears at 467 cm$^{-1}$ below 200 K and slightly increases in frequency at lower temperatures. In addition, Raman spectrum of CuCrO$_2$ at 8 K shows a broad band centered at 550 cm$^{-1}$. This mode could be due to magnon modes reflecting the proper screw spin structure below $T_{N2}$ = 23.6 K in CuCrO$_2$. }
		\label{ccocfot}
		\end{center}
		\end{figure}

Unpolarized Raman spectra of CuFeO$_2$ obtained between 290 K and 5 K are presented in Fig.~\ref{ccocfot}a. Over this temperature range, no considerable change is observed. In particular, no splitting of the $E_g$ mode below $T_{N1}$ is noticeable despite the $R\overline{3}m \rightharpoonup C2/m $ structural transition at $T_{N1}$ \cite{quirion2008, terada2006,ye2006}. We attribute this discrepancy to weak resonance with the He-Ne excitation line, which results in the weak intensity of the E$_g$ mode and makes it difficult to resolve any possible splitting. Another possibility is that the temperature of the sample remains above $T_{N1}$ even with a beam power of 4000 W/cm$^2$. 
		
In the case of CuCrO$_2$, unpolarized Raman spectra shown in Fig.~\ref{ccocfot}b display noticeable differences as the temperature is decreased from room temperature down to 8 K. With a close look at the E$_g$ mode at 458 cm$^{-1}$, one can observe that its tale becomes broader on the right hand side starting at 200 K. With further cooling, an additional mode is easily distinguished and its frequency increases to 470 cm$^{-1}$ at 8 K. This mode is also observed with parallel and cross polarization configurations (not shown). It should be noted that neutron diffraction measurements \cite{terada2006, ye2006}, magnetostriction measurements \cite{kkimura2009} and sound velocity measurements (Fig.~\ref{C66}) do not show any anomaly that could be associated with a structural deformation in the temperature range around 200 K. This mode could have an origin similar to that of the additional modes observed between room temperature and 8 K (Fig.~\ref{ccocfot}b). Moreover, the spectrum of CuCrO$_2$ at 8 K (Fig.~\ref{ccocfot}b) deserves some attention. Unlike the spectra at other temperatures, it develops a broad background feature centered at $\sim$550 cm$^{-1}$. This broad band, which can also be observed using parallel and cross polarizations (not shown), might be associated with magnon modes owed to a proper screw ordering observed below $T_{N2}$ \cite{poienar2010,kajimoto2010}. Finally, although there is some evidence for a structural deformation at $T_{N1}$ in CuCrO$_2$ \cite{kkimura2009}, no additional Raman modes are observed below this temperature. Although local heating due to incident beam power (4000 W/cm$^2$) is possible, the broad band observed in the spectrum at 8 K (Fig.~\ref{ccocfot}b)) clearly shows that the temperature is below $T_{N1}$.  Linewidths of Raman modes normally narrow down with decreasing temperature. For example, in BiFeO$_3$, which undergoes a structural transition at the Curie temperature $T_c$ = 1100 K, Fukumura et. al. \cite{fukumura2007} observed only 7 Raman modes at room temperature due to broadening of the modes. All 13 Raman modes were observed only at 4 K much below $T_c$ \cite{fukumura2007}. For CuCrO$_2$, even lower temperatures and a lower incident beam power might be required for the observation of additional modes.
 
\begin{figure}[htbp]
\begin{center}
		\includegraphics[width=12.0cm]{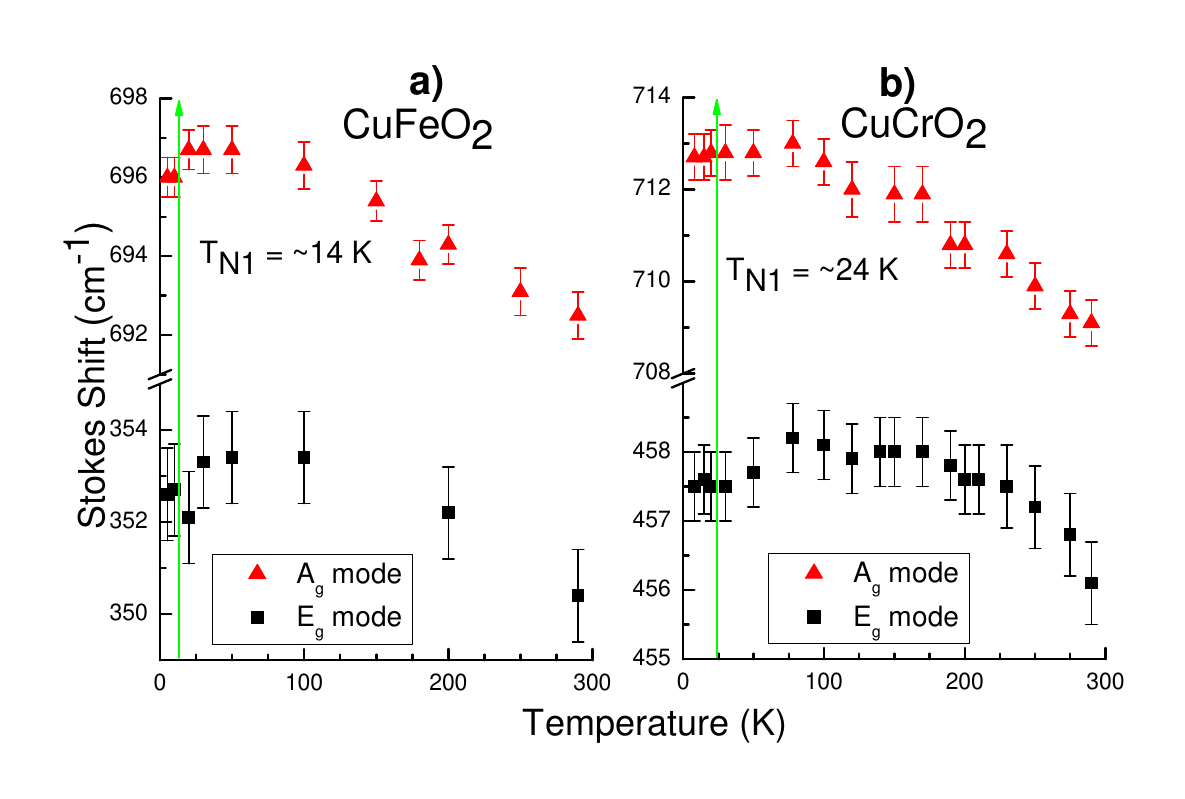}
		%\end{center}
		\caption{Temperature dependencies of the frequencies of the Raman modes in \textbf{a)} CuFeO$_2$ and \textbf{b)} CuCrO$_2$ between 295 K and 5 K could be associated with lattice contraction and anharmonic phonon-phonon interactions. Therefore, the E$_g$ modes in these compounds are not associated with the order parameters of the pseudoproper ferroelastic transitions at $T_{N1}$ in CuFeO$_2$ and CuCrO$_2$. }
		\label{agegcfocco}
		\end{center}
		\end{figure}

Temperature variations of the frequencies of the Raman modes in CuFeO$_2$ and CuCrO$_2$ are presented in Fig.~\ref{agegcfocco}. As shown in Fig.~\ref{agegcfocco}a, the frequencies of both modes in CuFeO$_2$ increase almost linearly down to 50 K with no significant variation below $T_{N1}$ = 14 K. Similarly, in the case of CuCrO$_2$ (Fig.~\ref{agegcfocco}b), the mode frequencies increase between 290 K and 80 K. While the A$_g$ mode frequency remains constant between 80 K and 8 K, the frequency of the E$_g$ mode seems to decrease slightly in this range. The additional modes (Fig.~\ref{ccocfot}) observed in both compounds behave similarly with temperature (not shown).

As discussed earlier, neutron \cite{ye2006} and x-ray \cite{terada2006} diffraction measurements on CuFeO$_2$ reveal an $R\overline{3}m \rightharpoonup C2/m $ structural transition while magnetostriction measurements on CuCrO$_2$ show evidence for crystal symmetry lowering. In accordance with these results, sound velocity measurements on
 CuFeO$_2$ \cite{quirion2008} and CuCrO$_2$ (see Fig.~\ref{C66}) indicate that both compounds undergo an $R\overline{3}m \rightharpoonup C2/m $ pseudoproper ferroelastic transition at $T_{N1}$. According to the group theory \cite{tinkhan}, one possible scenario is that an E$_g$-symmetric optic mode is associated with the order parameter \cite{quirion2008}. In this case, the excess Gibbs free energy $G_e$ can be written as
\begin{equation}
G_e = \frac{1}{2}m{w_{o}}^2 u^{2} + \frac{1}{2} b u^4 + \frac{1}{2} C e^{2}_{s} + \gamma e_{s} u,
\label{landauson}
\end{equation}
where
\begin{equation}
m{w_{o}}^2 = a(T - T_o) = A.
\label{A}
\end{equation}
In the above equations, $m$ is the reduced mass, $w_{o}$ is the uncoupled frequency of the soft E$_g$ mode, while $a$ and $b$ are temperature independent constants. The first two terms in Eq.~\ref{landauson} are due to the Landau expansion of the order parameter $u$, which corresponds to normal coordinate vibrations associated with the soft E$_g$ optic mode. Thus, the first term $\frac{1}{2}m{w_{o}}^2 u^{2}$ corresponds to the harmonic oscillator energy and is the only temperature dependent term. The term $\frac{1}{2} C e^{2}_{s}$ is the elastic energy of the soft acoustic mode associated with the strain component $e_s$. For simplification, only one elastic constant $C$ is considered (See Ref.~\cite{quirion2008} for complete elastic energy). Finally, the bilinear coupling term $\gamma e_{s} u$ in Eq. \ref{landauson}, with $\gamma$ representing the coupling coefficient, is necessary in order to account for the softening of the acoustic modes observed in CuFeO$_2$ \cite{quirion2008} and CuCrO$_2$ (see Fig.~\ref{C66}). Minimizing $G_e$ with respect to $u$ and $e_s$, one obtains
\begin{equation}
e_s = -\frac{u \gamma}{C}
\label{sstrain}
\end{equation}
and
\begin{equation}
u = \frac{\sqrt{a(T_{o}-T + \frac{\gamma^{2}}{C})}}{\sqrt{B}},
\label{OPlandauTc}
\end{equation}
showing that the bilinear coupling term ($\gamma e_s u$) renormalizes the uncoupled transition temperature $T_o$ to
\begin{equation}
T_{N1} = T_o + \frac{{\gamma}^2}{aC}.
\label{ToTc}
\end{equation}
Finally, the frequency $\omega$ of the soft optical mode can be obtained using \cite{david}
\begin{equation}
 \omega^2 = \frac{1}{m} \frac{{\partial}^2 G_e}{\partial u^2},
\label{w2}
\end{equation}
which yields
\begin{equation}
\omega^2 = \frac{a}{m}(T-T_{N1})+\frac{{\gamma}^2}{C m} \hspace{1.4pc} (\mathrm{T}\ > \ T_c)
\label{w2highTc}
\end{equation}
and
\begin{equation}
\omega^2 = \frac{-2a}{m}(T-T_{N1})+ \frac{{\gamma}^2}{C m} \hspace{1.4pc} (\mathrm{T}\ < \ T_c).
\label{w2lowTc}
\end{equation}
According to Eqs. \ref{w2highTc} and \ref{w2lowTc}, the frequency square of the soft optical mode should vary linearly with temperature with a slope change at T$_{N1}$ as observed in some pseudoproper ferroelastic compounds \cite{oktay, quirion2009RLHS, sno2, bivo4}. According to our Raman measurements, the temperature dependence of the E$_g$ symmetry modes cannot be associated with that of a soft optic mode. Thus, the temperature behavior of all modes is rather attributed to thermal contraction and anharmonic phonon-phonon interactions, in agreement with the analyses of Pavunny et. al. \cite{pavunny2010} for their Raman study of CuFeO$_2$ between 400 K and 80 K. Our conclusion is that none of the Raman modes in CuFeO$_{2}$ and CuCrO$_2$ can account for the pseudoproper ferroelastic transitions observed at $T_{N1}$ = 14 K in CuFeO$_2$ and at $T_{N1}$ = 24.3 K in CuCrO$_2$. While these results do not refute the pseudoproper ferroelastic transitions in CuFeO$_2$ \cite{quirion2008} and CuCrO$_2$ (see Fig.~\ref{C66}), they leave the driving mechanisms unresolved. 

\section{Conclusions \label{conclusions}}

Polarized Raman scattering measurements were performed on delafossite magnetoelectric CuFeO$_2$ in order to determine the true nature of the order parameter associated with pseudoproper ferroelastic transition observed in CuFeO$_2$ by means of sound velocity measurements \cite{quirion2008}. As preliminary sound velocity measurements on the isostructural compound CuCrO$_2$ show similar elastic softening at the antiferromagnetic transition (see Fig.~\ref{C66}), the Raman measurements were also performed on CuCrO$_2$ for comparison. 
 
Apart from the vibrational modes in CuFeO$_2$ and CuCrO$_2$ with A$_g$ and E$_g$ symmetries, one additional mode in CuFeO$_2$ and seven additional modes in CuCrO$_2$ are observed at room temperature. Below 200 K, another mode in CuCrO$_2$ with a frequency close to that of the E$_g$ mode appears and persists to lower temperatures. The additional modes observed in both compounds are possibly associated with either relaxation of Raman selection rules or crystal field excitations. More interestingly, the spectrum of CuCrO$_2$ at 8 K shows a broad band centered at 550 cm$^{-1}$ attributed to the proper screw ordering below $T_{N2}$ in CuCrO$_2$. Furthermore, for both compounds the frequency of all modes increase with decreasing temperature, with no significant variation at the phase transitions. The observed temperature behavior is thus attributed to thermal contraction and anharmonic phonon-phonon interactions. Therefore, these results show that the $E_g$ symmetry Raman modes in CuFeO$_2$ and CuCrO$_2$ do not induce the transitions observed at $T_{N1}$, leading to the necessity of further search for the true origin of the order parameters associated with the pseudoproper ferroelastic transitions observed at $T_{N1}$ in both compounds.

\section{Acknowledgements}
We would like to thank Oleg Petrenko and Serge Jandl for fruitful discussions. This work was supported by grants from the Natural Science and Engineering Research Council of Canada (NSERC) as well as from the Canada Foundation for Innovation (CFI).

\end{document}